\documentstyle[espcrc2,fleqn,twoside,epsfig]{article}

\title{Effective Macroscopic Response of a Composite with
Small Deviations from Periodicity: Application to Colloidal Crystals}

\author{Sergey V. Barabash, David Stroud
\address{
Department of Physics,
The Ohio State University, Columbus, Ohio 43210}
}

\begin{document}
\def\etal{{\em et al}}
\def\eps{\epsilon}
\def\veps{\varepsilon}
\def\bk{ {\bf k} }
\def\bq{ {\bf q} }
\def\bkp{ {\bf k^\prime} }
\def\bRa{ {\bf R_a} }
\def\bRb{ {\bf R_b} }
\def\bDR{ \Delta{\bf R} }
\def\bra{ {\bf r_a} }
\def\brb{ {\bf r_b} }
\newcommand{\beq}{\begin{equation}}
\newcommand{\unbeq}{\end{equation}}
\newcommand{\eeq}{\end{equation}}
\newcommand{\ba}{\begin{eqnarray}}
\newcommand{\unba}{\end{eqnarray}}
\newcommand{\ea}{\end{eqnarray}}




\begin{abstract}
\vspace{2pc}

Using the spectral approach, we analyze the effective
properties of a composite which deviates slightly from
periodicity.
We find that, when the inclusions are randomly displaced 
from their equilibrium positions, the sharp resonances seen in 
the periodic case are broadened, and an additional branch cut
appears.   
We use these results to analyze
the effective dielectric constant of a colloidal crystal.  

\vspace{5pc}
\end{abstract}
\maketitle



The spectral approach\cite{Bergman} has proven useful in
analyzing the effective dielectric constant of a composite material.
For example, it has led to the derivation of exact 
bounds on such effective properties\cite{Milton,SSP}, as well as
to approximate calculations of effective nonlinear 
properties\cite{Sheng98};
and it can be generalized to polycrystalline 
materials\cite{BarabashStroud99}.  However, it has proven difficult
to apply to the spectra of real systems, with
the exception of a few simple cases,
such as a simple cubic lattice of identical 
spheres\cite{Bergman,BergmanJPC79}.
%

In this paper, we achieve some progress in this direction 
by calculating the effect of weak disorder on a 
periodic arrangement of identical inclusions.
Our results 
may prove useful in describing the effective properties
of real systems such as colloidal crystals.      

First we review the periodic case.
Consider a system of identical inclusions (labeled by the
index $a$). 
Let the spectrum of each inclusion be described by the poles
$\{s_{\alpha a}\}=\{s_\alpha\}$ with the corresponding amplitudes
$\{M_{\alpha a}\}=\{M_\alpha\}$, and let
$Q_{\alpha a,\beta b}$ be the overlap integrals between the 
inclusions (see Ref. \cite{Bergman} for further
definitions).  
The poles of the system are then given by the solutions of
\beq
(s-s_{\alpha a}) A_{\alpha a}= 
\sum_{\beta, b\neq a} 
	Q_{\alpha a,\beta b} A_{\beta b},
\label{eq:poles}
\eeq
where the amplitude of the pole $s^{(i)}$is
\beq
M^{(i)}=\sum_{\alpha a}A_{\alpha a}^{(i)} M_{\alpha a} /
\sqrt{\sum_{\alpha a} |A_{\alpha a}^{(i)}|^2}.
\label{def:M}
\eeq
In the purely periodic case, Bloch theorem implies that
$A_{\alpha a}= A_\alpha({\bf k}) e^{i{\bf k\cdot R_a}}$,
where ${\bf R_a}$ is the position of the inclusion $a$. 
Since $Q_{\alpha a,\beta b}=Q_{\alpha\beta}(\bRb-\bRa)$, it
then follows\cite{Bergman} that
the spectrum is a system of ``bands'' given by 
\beq
(s-s_{\alpha}) A_{\alpha}(\bk)= 
	\sum_{\beta} \tilde Q_{\alpha,\beta}(\bk) A_\beta(\bk),
\label{eq:bands}
\eeq
where
\beq
\tilde Q_{\alpha\beta}(\bk)=
\sum_{b\neq a}Q_{\alpha a,\beta b}
	e^{-i\bk\cdot (\bRa-{\bf R_b}) }.
\label{eq:tildeQ}
\eeq
For the periodic case, the entire amplitude within each band  
is concentrated in the $\bk=0$ pole: 
\beq
M^{(i)}(\bk)= \frac{
	\delta_{\bk,0} \sqrt{N} 
	\sum_{\alpha}A^{(i)}_\alpha(0)M_\alpha
	}{  \sqrt{\sum_\alpha|A^{(i)}_\alpha(0)|^2}
	}, 
\label{M0}
\eeq
where $N\to\infty$ is the number of inclusions in the system.

We now introduce disorder into the system by 
displacing the inclusions to positions
\beq
\bra= \bRa + \sum_\bk {\bf U_k} e^{i \bk\cdot\bRa}.
\label{eq:positions}
\eeq
Eq.\ (\ref{eq:positions}) could, for example, describe a system of 
phonons in a colloidal crystal. Each ${\bf U_k}$ defines the amplitude
and polarization of a phonon with wave vector $\bk$. We will assume
that all nonzero ${\bf U_k}$'s are of the same order of magnitude:
$|{\bf U_k}|\sim U_{ph}$, and treat $U_{ph}$ as a small parameter.
To find the effect of such a ``phononic'' perturbation 
on the spectrum of the system, we express $A_{\alpha a}$ as
\beq
A_{\alpha a}=\sum_{\bf k} A_\alpha({\bf k}) e^{i{\bf k\cdot R_a}},
\label{eq:Bloch}
\eeq
where ${\bf k}$ is confined to the first Brillouin zone.
Substituting (\ref{eq:Bloch}) into (\ref{eq:poles}), we find that 
Eqs.\ (\ref{eq:bands}-\ref{eq:tildeQ}) should now be generalized to
\beq
(s-s_{\alpha}) A_{\alpha}(\bk)= 
  \sum_{\beta\bkp} \tilde Q_{\alpha,\beta}(\bk,\bkp) A_\beta(\bkp),
\label{eq:newbands}
\eeq
\beq
\tilde Q_{\alpha,\beta}(\bk,\bkp)=
\frac{1}{N}
\sum_a\sum_{b\neq a} Q_{\alpha a,\beta b}\,
	e^{i(\bkp\cdot\bRb - \bk\cdot\bRa) }.
\label{def:Qkk}
\eeq
If we restrict our consideration to the correction of the lowest order
in $U_{ph}$, we can explicitly write down the approximate expression
for the position-dependent overlap integrals $Q_{\alpha a,\beta b}$:
\ba
Q_{\alpha a,\beta b} &=&
     Q_{\alpha,\beta}(\bRb-\bRa) + \delta Q^{(1)}_{\alpha a,\beta b};
\nonumber \\
\delta Q^{(1)}_{\alpha a,\beta b} &\equiv&   \sum_\bk {\bf U_k} \cdot\left( 
		\nabla Q_{\alpha,\beta}(\bRb-\bRa) 
	\right) \nonumber \\
	&&\times\left( e^{i \bk\cdot\bRb} - e^{i \bk\cdot\bRa}\right).
\label{eq:Q1}
\ea
Plugging in this expression for $\delta Q^{(1)}_{\alpha a,\beta b}$
back into Eq.\ (\ref{def:Qkk}), we obtain:
\beq 
\tilde Q_{\alpha,\beta}(\bk,\bkp) 
   = \delta_{\bk,\bkp} \tilde Q_{\alpha\beta}(\bk) 
  	+ \,\,  {\bf U_{k-k^\prime}} \cdot 
		{\bf \vec Q}_{\alpha,\beta}(\bk,\bkp),
\eeq
where
\beq
{\bf \vec Q}_{\alpha,\beta}(\bk,\bkp) =
	 (\widetilde{ \nabla Q}_{\alpha,\beta})_\bk 
	-(\widetilde{ \nabla Q}_{\alpha,\beta})_\bkp
\label{eq:vecQ}
\eeq
and
\beq
(\widetilde{ \nabla Q}_{\alpha,\beta})_\bk =
	\sum_{\bDR\neq 0} \left[ \nabla Q_{\alpha,\beta}(\bDR) \right]
	\times e^{i\bk\bDR},
\label{eq:tildeNablaQ}
\eeq
with $\bDR\equiv\bRb-\bRa$. Note that because the Fourier transforms
(\ref{eq:tildeQ}) and (\ref{eq:tildeNablaQ}) involve only discrete sums
rather than continuous integrals, one cannot assume that 
$(\widetilde{\nabla Q}_{\alpha,\beta})_\bk = \bk\tilde Q_{\alpha,\beta}(\bk)$.
Instead, both (\ref{eq:tildeQ}) and (\ref{eq:tildeNablaQ}) should be evaluated
directly from the known functional form of $Q_{\alpha,\beta}(\bDR)$.

The Eq.(\ref{eq:newbands}) now takes the form
\ba
&&\!\!\! \!\!\! \!\!\! \!\!\! (s-s_{\alpha}) A_{\alpha}(\bk)= 
  \sum_\beta \left\{ \tilde Q_{\alpha,\beta}(\bk)A_\beta(\bk) \right. 
\label{eq:linearbands}
\\
&&	+ \sum_\bkp \left. {\bf U_{k-k^\prime}}\cdot 
		{\bf \vec Q}_{\alpha,\beta}(\bk,\bkp)
		 A_\beta(\bkp)\right\}.
\nonumber
\ea
The solutions of this equation give the new positions of the poles and
the corresponding eigenvectors. Note that unlike Eq. (\ref{eq:bands}),
which could be solved for each value of $\bk$ independently, a single solution
of (\ref{eq:linearbands}) generally involves an infinite number of amplitudes
$A_\alpha(\bkp)$.

We now explicitly find the solutions of (\ref{eq:linearbands}),
restricting ourselves to the case when 
in the purely periodic system the pole at 
$s_0^{(i)}(\bk_0)$ is not degenerate.
We can
expand the corresponding eigenvector $A^{(i,\bk_0)}_\alpha(\bk)$
of Eq.(\ref{eq:linearbands})
in terms of the unperturbed eigenvectors ${A_0}^{(j)}_\alpha(\bk)$ by
writing
\beq
A^{(i,\bk_0)}_\alpha(\bk)=
 {A_0}^{(i)}_\alpha(\bk) \delta_{\bk,\bk_0}
	+ \sum_{j} 
			c^{(i,\bk_0)}_{j,\bk}{A_0}^{(j)}_\alpha(\bk),
\eeq
where the factors $c^{(i,\bk_0)}_{j,\bk}$ are of the first order in 
phononic amplitudes.
This can be done because\cite{Bergman}
$\tilde Q^*_{\alpha,\beta}(\bk) = \tilde Q_{\beta,\alpha}(\bk)$,
and thus the different solutions ${A_0}^{(j)}_\alpha(\bk)$
of (\ref{eq:bands}) do form a complete set at each given 
value of $\bk$;
we also assume that
$\sum_\alpha {A_0}^{(i)}_\alpha(\bk)^* {A_0}^{(j)}_\alpha(\bk) = \delta_{i,j}$
Using the latter property and noting that
\ba
\sum_\beta&& \!\!\! \!\!\! \!\!\!\!\!\!
	 \left[ 
	\tilde Q_{\alpha,\beta}(\bk) - s_\alpha \delta_{\alpha,\beta}
	\right] 
	{A_0}^{(i)}_\beta(\bk) \nonumber\\
 &&\,\,\,\,\, \,\,\,\,\, \,\,\,\,\, = \,\, 
	s_0^{(i)}(\bk) {A_0}^{(i)}_\alpha(\bk)
\ea
[cf. eq.\ (\ref{eq:bands})]
it is straightforward to show that the first order correction
to the eigenvector is given by
\ba
&&\!\!\! \!\!\! \!\!\! \!\!\!
	c^{(i,\bk_0)}_{j,\bk} =   
\label{eq:c}
\\
&&
 \frac{	\sum\limits_{\alpha,\beta}
		A_{0_\alpha}^{(j)}(\bk)^*
		{\bf U_{k-k_0}} \cdot
		{\bf \vec Q}_{\alpha,\beta}(\bk,\bk_0)
		A_{0_\beta}^{(i)}(\bk_0)
				}{
		s_0^{(i)}(\bk_0) -s_0^{(j)}(\bk) 
				}\nonumber
\ea
for $\bk\neq\bk_0$.
For $\bk=\bk_0$, $c^{(i,\bk_0)}_{j,\bk_0}=0$,
reflecting the fact that the diagonal term of the perturbation is zero:
${\bf \vec Q}_{\alpha,\beta}(\bk,\bk) =0$.
For the same reason the position of the non-degenerate pole $s_0^{(i)}(\bk_0)$ 
remains unchanged in the first order.

Next, we analyze how the amplitude of the non-degenerate pole is affected by the
disorder. From (\ref{eq:Bloch}) and (\ref{def:M}) we get
\beq
M^{(i,\bk_0)} = \frac { 
	\sum_\alpha N M_\alpha 
	A^{(i,\bk_0)}_\alpha (0)
			}{\sqrt{
	N \sum_\alpha \sum_\bk |A^{(i,\bk_0)}_\alpha (\bk)|^2
			}}.
\label{eq:Mgeneral}
\eeq
Up to the terms of the second order, the sum in the denominator equals
$\sum_\alpha |{A_0}^{(i)}_\alpha(0)|^2$ which is simply unity by our
our choice of orthonormal set of ${A_0}^{(i)}_\alpha(\bk)$'s.
Thus, in the non-degenerate case the amplitude of the pole at 
$\bk_0=0$ remains unchanged in the first order.
The amplitudes at other poles are:
\beq
M^{(i,\bk)} = 
	\sqrt{N}\sum_\alpha M_\alpha
	\sum_{j} 
	c^{(i,\bk)}_{j,0} A_{0_\alpha}^{(j)}(0),
\label{result:M}
\eeq
where
\beq
c^{(i,\bk)}_{j,0} =   
 \frac{	\sum\limits_{\alpha,\beta}
		A_{0_\alpha}^{(j)}(0)^*
		{\bf \vec Q}_{\alpha,\beta}(0,\bk)
		\cdot
		{\bf U_{-k}} 
		A_{0_\beta}^{(i)}(\bk)
				}{
		s_0^{(i)}(\bk) -s_0^{(j)}(0) 
				}
\label{eq:c0s}
\eeq
Thus, for each nonzero Fourier component ${\bf U_{k}}$ of the disorder,
the poles $s^{(i,-\bk)}$ in each band pick up nonzero amplitude
$M^{(i,-\bk)}\sim |{\bf U_{k}}| \sim U_{ph}$.

What happens in the vicinity of the unperturbed poles ($\bk=0$)?
Because $Q_{\alpha, \beta}(\bDR)\to 0$ as $\bDR\to\infty$, it follows
from eq.\ (\ref{eq:vecQ}) that
${\bf \vec Q}_{\alpha,\beta}(0,\bk\to 0) \to 0$.
This, however, does not directly relate to the small-$\bk$ behavior
of $M^{(i,\bk)}$, since one also has
$s_0^{(i)}(\bk\to 0) \to s_0^{(j)}(0)$ at least for
$i=j$. It is natural to expect that $s_0^{(i)}(0)$ is either at the top
or at the bottom of the $i$-th band, in which case
$s_0^{(i)}(\bk) - s_0^{(i)}(0) \sim a k^2$
for $\bk\to 0$, whereas the ${\bf \vec Q}$'s are likely to follow
${\bf \vec Q}_{\alpha,\beta}(0,\bk) \sim \bk$ [which would be the
case if (\ref{eq:tildeQ}) and (\ref{eq:tildeNablaQ}) were continuum
Fourier transforms]. 
In this case, eq.\ (\ref{eq:c0s}) predicts that
$c^{(i,\bk)}_{j,0} \to \infty$ as $\bk\to 0$,
which means that the expression 
for the first-order non-degenerate case cannot be used under these conditions.
Instead, for sufficiently small values of $\bk$ a degenerate theory should be
applied. Namely, for the states with $|\bk| < k_c$,
where $k_c$ is chosen so that 
$ s_0^{(i)}(\bk_c) -s_0^{(i)}(0) \ll
	\sum\limits_{\alpha,\beta}
		A_{0_\alpha}^{(j)}(0)
		{\bf \vec Q}_{\alpha,\beta}(0,\bk_c)
		\cdot
		{\bf U_{-k_c}} 
		A_{0_\beta}^{(i)}(\bk_c)
$,
we can make the approximation $s_0^{(i)}(\bk) \approx s_0^{(i)}(0)$.
In the first order this will not change the positions of the poles,
because $k_c\sim U_{ph}$ and thus
${\bf \vec Q}_{\alpha,\beta}(0,\bk_c)\cdot {\bf U_{k_c}} \sim {U_{ph}}^2$. 
However, all the eigenvectors of the degenerate problem would contain
$A_{0_\beta}^{(i)}(0)$ with finite coefficients. Thus, 
the unperturbed weight
$|M^{(i)}(0)|^2$ of the $\bk=0$ pole [see eq.\ (\ref{M0})] 
will become distributed between all the poles
$s^{(i)}(\bk)$ that correspond to $\bk<\bk_c$ 
(and such that ${\bf U_{-k}}\neq 0$).

\begin{figure}[hbt]
\epsfig{file=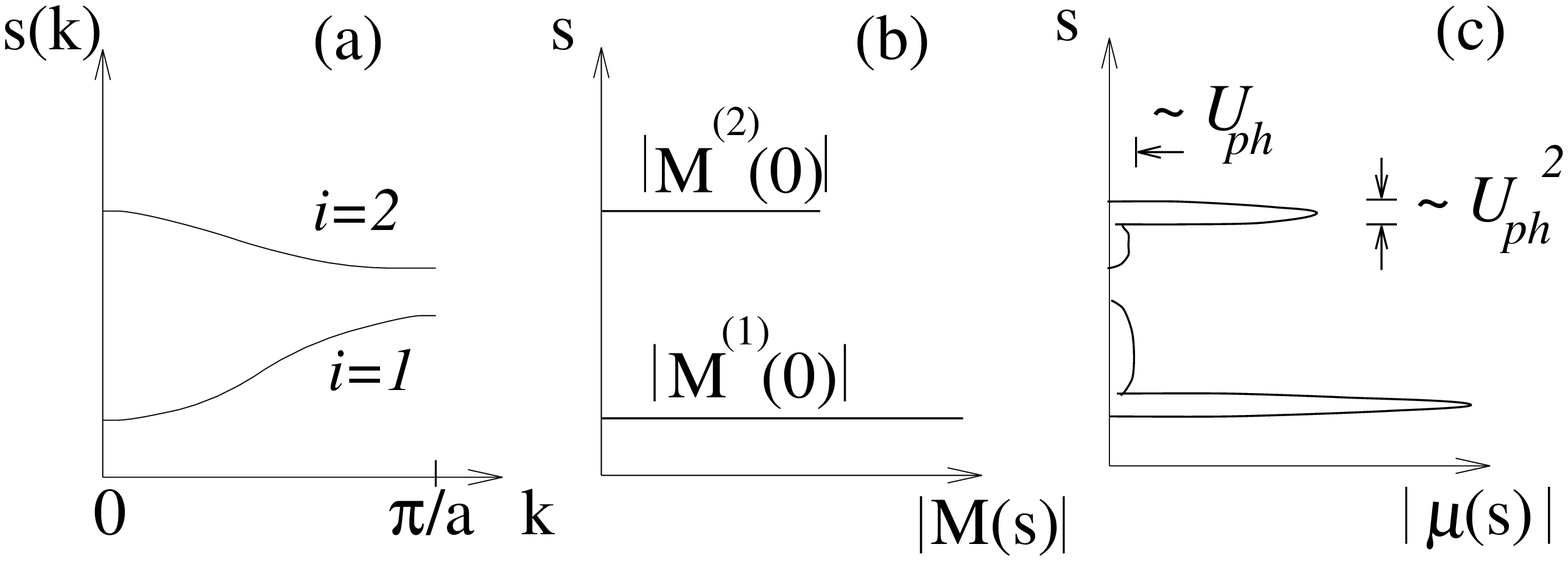, width=3in}
\caption{Schematic of (a) the ``band structure'' for a periodic 
composite (two ``bands'' are shown); and the
corresponding pole spectrum for (b) a purely periodic, and 
(c) a slightly disordered system 
(with characteristic ``phononic'' amplitude $U_{ph}$).
}
\label{fig1}
\end{figure}

Consider, for example, an infinite sample in which each inclusion is
randomly displaced from its original position. The sum in 
Eq.\ (\ref{eq:positions}) should then be replaced by an integral
over the first Brillouin zone, so that there will be a
continuous density of poles with nonzero weight.
The corresponding spectral function is written 
\beq
F(s)=
\int_0^1 	\frac{|\mu(s')|^2 }{s-s'} ds',
\label{eq:integralF}
\eeq
where $|\mu(s)|^2$ is interpreted as a density of pole weight.
In the purely periodic case (or for a finite system),
\beq
|\mu_0(s)|^2=\sum_i |M^{(i)}(0)|^2 \delta\left(s-s^{(i)}(0)\right),
\label{eq:mu0}
\eeq
so that $F(s)$ 
is non-analytic only at 
the simple poles $s^{(i)}(0)$\cite{Bergman}.
In the presence of disorder, however,
\beq
\mu(s) =\mu_{peak}(s) + \mu_{band}(s),
\eeq
where $\mu_{peak}(s)$ represents the amplitudes of
the discrete poles $s^{(i)}(\bk)$ which are
degenerate with $s^{(i)}(0)$, while
$\mu_{band}(s)$ gives the amplitude density
for the rest of the band.
From Eqs.\ (\ref{result:M}-\ref{eq:c0s}), 
$\mu_{band}(s)\sim U_{ph}$.
For the typical case discussed in the previous paragraph,
the width of the peak part can be estimated by noting that
$s_0^{(i)}(k_c) - s_0^{(i)}(0) \sim a k_c^2 \sim U_{ph}^2$.
Therefore, we expect each $\delta$-function entering $\mu_0(s)$  
to acquire a half-width proportional to $U_{ph}^2$.
This typical situation is illustrated in Fig.~\ref{fig1}.

For $s$ sufficiently different from $s_0^{(i)}(0)$, 
$\mu_{peak}(s)$ can probably be
replaced by a set of $\delta$-functions:
\beq
F(s)\approx
\!\!\!\!\!\!
 \sum_{ s'\in \{ s^{(i)}(0) \} } \!\!\!
 	\frac{ |M^{(i)}(0) |^2}{s-s'}
+ \int_0^1 
	\frac{|\mu_{band}(s')|^2 }{s-s'} ds'
\label{eq:integralSumF}
\eeq
The second term describes 
the part of the spectral function which cannot be characterized
purely by simple poles. 
Any such function can be described by
a branch cut along the segment $[0,1)$ of the real axis, 
as can be seen by writing the Cauchy formula for the contour encircling
the branch cut infinitesimally below the real axis.
For example,
the effective medium approximation (EMA)\cite{SSP} gives
such a branch cut.
While branch cut in the present case
would certainly differ from the EMA one,
it is quite remarkable that such a cut would appear even in a
weakly disordered system.


A possible application of this work could be a colloidal crystal
at some finite temperature $T$.
The total contribution to the spectral function
from the integral in (\ref{eq:integralSumF}) involves
$|\mu_{band}(s')|^2$, which 
is determined by $\sum_\bq |{\bf U_q}|^2$. 
In a conventional crystal, the analogous sum increases linearly
in $T$ at high $T$ and approaches constant value as $T\to 0$, 
and similar behavior should be observed in a
colloidal crystal.  

We thus suggest that in a colloidal crystal,
at $T$ such that first order corrections are adequate,
the spectrum should consist of two parts: (a) electrostatic
resonances at the same ratios
$\eps_1/\eps_2$ predicted by the theory for the periodic 
case [see \cite{BergmanJPC79} for explicit expressions for
spherical inclusions], but slightly broadened by a half-width
proportional to $T^2$; and (b) a continuous contribution from
a branch cut introduced by the disorder.
The latter contribution should be most prominent 
at negative values of $\eps_1/\eps_2$ away from
the original resonances.
It would be of great interest if such a spectrum could be
detected in a real material, e. g., in a suspension of metal
spheres in a dielectric host.


This work has been supported by NSF Grant
DMR01-04987, and by the U.-S./Israel Binational Science Foundation.
We thank Prof. David Bergman and Oleg Lunin for valuable conversations.


\begin{thebibliography}{00}

\bibitem{Bergman}
D. J. Bergman
Phys. Rep. {\bf 43}, 377 (1978);
Phys. Rev. B {\bf 19}, 2359 (1979).

\bibitem{Milton} R. C. McPhedran and G. W. Milton, 
Appl. Phys. A {\bf 26}, 207 (1981); 
G. W. Milton, J. Appl. Phys. {\bf 52}, 5294 (1981).

\bibitem{SSP}
For a review see D. J. Bergman and D. Stroud, 
in {\it Solid State Physics}, edited by H. Ehrenreich and D. Turnbull 
(Academic, New York, 1992), Vol. 46, pp.\ 178-320.


\bibitem{Sheng98}
H. Ma, Xiao, P. Sheng,
J. Opt. Soc. Am. B {\bf 15}, 1022 (1998).

\bibitem{BarabashStroud99} S. Barabash and D. Stroud,
J. Phys.: Condens. Matter {\bf 11}, 10323 (1999).

\bibitem{BergmanJPC79} D. J. Bergman, J. Phys. C {\bf 12}, 4947 (1979).




\end{thebibliography}
\end{document}